\documentclass[letterpaper,  
              ]{jacow}
\makeatletter%
	\ifboolexpr{bool{xetex}}
	 {\renewcommand{\Gin@extensions}{.pdf,%
	                    .png,.jpg,.bmp,.pict,.tif,.psd,.mac,.sga,.tga,.gif,%
	                    .eps,.ps,%
	                    }}{}
\makeatother

\ifboolexpr{bool{xetex} or bool{luatex}} 
 {}                                      
 {\usepackage[utf8]{inputenc}}           

\usepackage[USenglish]{babel}			 

\usepackage[final]{pdfpages}
\usepackage{multirow}
\usepackage{ragged2e}

\ifboolexpr{bool{jacowbiblatex}}%
 {%
  \addbibresource{jacow-test.bib}
  \addbibresource{biblatex-examples.bib}
 }{}
\begin{document}

\title{USING SLOPPY MODELS FOR CONSTRAINED EMITTANCE MINIMIZATION AT THE CORNELL ELECTRON STORAGE RING (CESR)}

\author{W. F. Bergan\thanks{wfb59@cornell.edu}, A. C. Bartnik, I. V. Bazarov, H. He, D. L. Rubin, and J. P. Sethna \\ Cornell University, Ithaca NY, USA}

\maketitle

\begin{abstract}
In order to minimize the emittance at the Cornell Electron Storage Ring (CESR), we measure and correct the orbit, dispersion, and transverse coupling of the beam. However, this method is limited by finite measurement resolution of the dispersion, and so a new procedure must be used to further reduce the emittance due to dispersion. In order to achieve this, we use a method based upon the theory of sloppy models. We use a model of the accelerator to create the Hessian matrix which encodes the effects of various corrector magnets on the vertical emittance. A singular value decomposition of this matrix yields the magnet combinations which have the greatest effect on the emittance. We can then adjust these magnet ``knobs'' sequentially in order to decrease the dispersion and the emittance. We present here comparisons of the effectiveness of this procedure in both experiment and simulation using a variety of CESR lattices. We also discuss techniques to minimize changes to parameters we have already corrected.
\end{abstract}

\section{Introduction}

Reduction of beam emittance in accelerators is important in enabling increased brightness for storage ring light sources as well as increased luminosity for damping rings for linear colliders. Accelerator physicists at the Cornell Electron Storage Ring (CESR) have been exploring ways to provide relatively fast emittance minimization by measuring and correcting orbit, dispersion, and coupling errors.\cite{cite:calibration} However, finite dispersion resolution ultimately limits the power of this method to eliminate the vertical dispersion as a source of emittance. In order to proceed further, we made use of the theory of sloppy models.\cite{cite:sethna}\cite{cite:thesis} This theory maintains that in certain systems with large numbers of free parameters, the parameters can be grouped as ordered ``eigenvectors'' such that the relative importance of successive eigenvectors decreases exponentially. If CESR is such a system, once we identify the eigenvectors, we will only need to work with the first few in order to achieve most of the desired results.

\begin{figure}[!htb]
   \centering
   \includegraphics*[width=225pt]{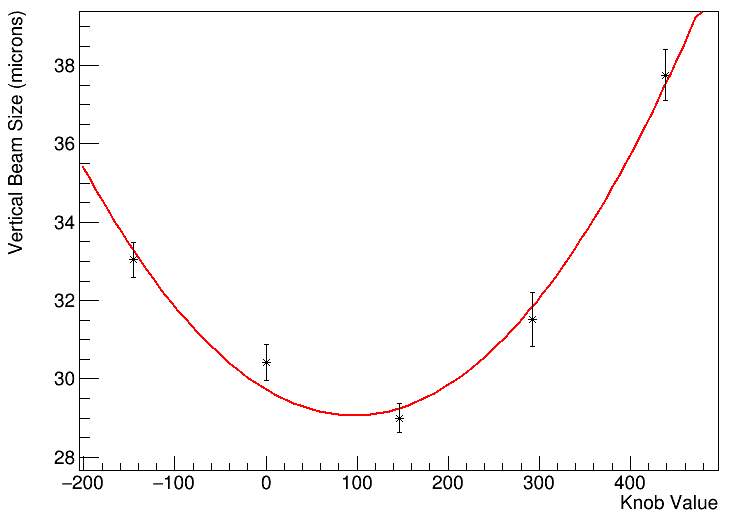}
   \caption{Beam size measured at a variety of knob values and fit to a parabola. Having performed this fit, we turn the knob to the parabola's minimum to reduce the emittance.}
   \label{fig:parabola}
\end{figure}

In light of this, we worked to identify combinations of magnets which have the largest impact on the vertical emittance. We then measured the dependence of the beam size on these knobs and tuned to minimize the emittance. We have also used a version of the robust conjugate direction search (RCDS) algorithm, as described by Huang et al, in order to enable computer-driven emittance minimization.\cite{cite:slac_1}\cite{cite:slac_2} Under this method, we sample the beam size at a variety of knob values in order to map out its functional form and fit it to a parabola. We then turn our knob to the value corresponding to the parabola's minimum and apply the same procedure to the subsequent knobs. An example of such a fit may be seen in Fig. \ref{fig:parabola}.

Prior results have shown that in minimizing the emittance, we often adversely affect other machine parameters.\cite{cite:napac} It is important for us to avoid these adverse effects as much as possible, and so we have found ways in which we may prevent these losses while still achieving most of the desired emittance reduction.

\section{Procedure}

\begin{figure}[!htb]
   \centering
   \includegraphics*[width=225pt]{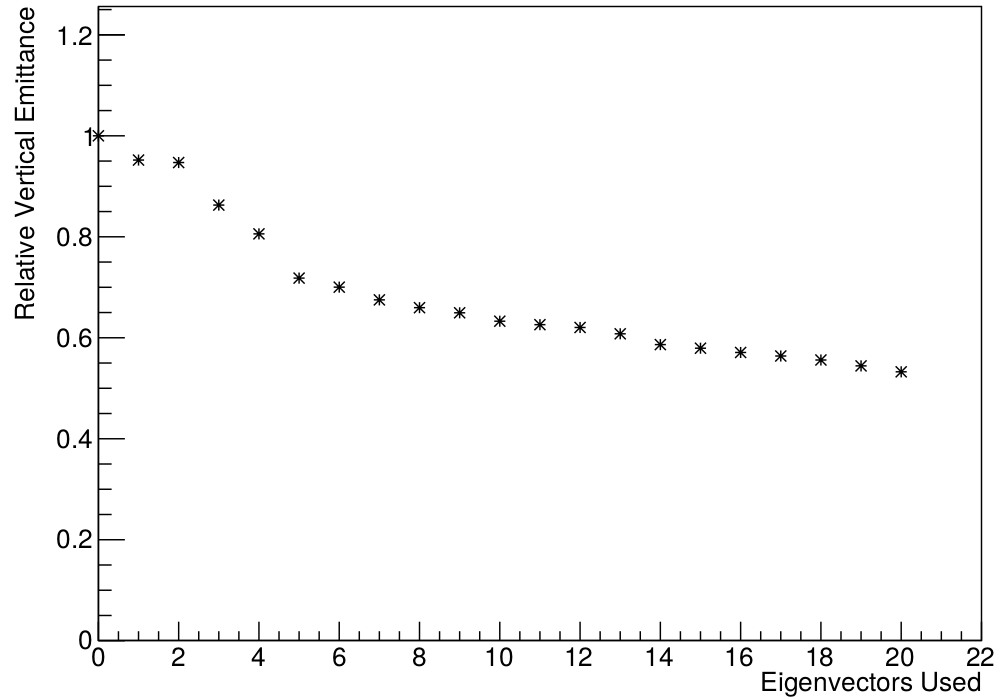}
   \caption{Change in emittance relative to starting emittance in simulation using knobs with no additional constraints. We have turned the knobs individually to minimize beam size, as we would in the real machine, and have averaged the results of 99 versions of the same 5 GeV lattice with random errors. Note that much of the emittance reduction is accomplished with the first few knobs, indicating that CESR qualifies as a ``sloppy'' system.}
   \label{fig:emittance_reduction}
\end{figure}

In order to identify the magnet combinations which have the largest effect on the vertical emittance, we used BMAD-based models of the CESR lattice.\cite{cite:bmad} We generated a lattice with errors introduced consistent with our known alignment tolerances and corrected it using our usual emittance-tuning procedure. We then used this lattice to compute the Hessian of the vertical emittance with respect to the various vertical kickers and skew quadrupoles, which we had found to be useful magnets for emittance-reduction.\footnote{The Hessian is not very sensitive to which errors we introduce to our lattice, but is different from the Hessian generated from the ideal lattice, and this method gave more effective knobs.} By taking the singular value decomposition of this matrix, we identified the combinations of magnets, the eigenvectors, having the largest effects on the emittance. We then tested these ``knobs'' on simulated CESR lattices with errors introduced consistent with known alignment tolerances, etc, and found that they are indeed effective. See, for example, Fig. \ref{fig:emittance_reduction}.

We have also been successful at turning these magnet combinations into real knobs which we have been able to use to reduce the emittance in CESR. In order to measure the emittance in the storage ring, we made use of either an x-ray beam-size monitor (xbsm) or visible-light beam size-monitor (vbsm), which measure the size of the beam at some fixed source point in the ring.\cite{cite:vbsm} On a 2 GeV lattice, we used the standard emittance-tuning techniques to reduce the beam size from 27 to 22 microns.\footnote{The knobs were computed separately for the 2 GeV and 5 GeV lattices. Note that, due to the different lattices, these values are different from the 5 GeV results which we will discuss later.} Using our first eight knobs while taking measurements with the xbsm, we were able to further reduce the beam size to 19 microns, demonstrating that these knobs are indeed effective at beam-size reduction.

\begin{figure}[!htb]
   \centering
   \includegraphics*[width=250pt]{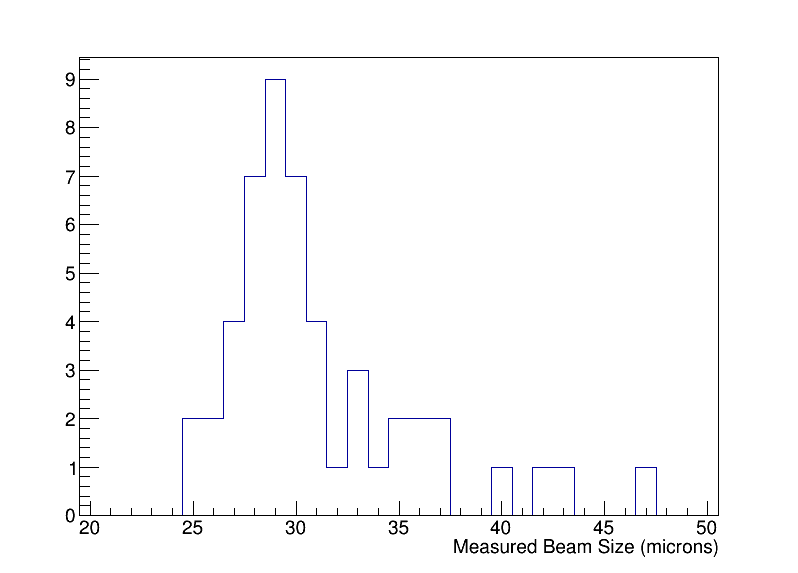}
   \caption{Distribution of 50 consecutive beam size measurements at a particular set of knob values. Note the long high-side tail.}
   \label{fig:high_side_tail}
\end{figure}

To perform the tests more efficiently, we used a variation on the RCDS algorithm. Using our knowledge of the relative strengths of the knobs, we were able to anticipate the sensitivity of the beam size to our various knobs, permitting us to choose appropriate step sizes for sampling the full parabolic structure. Also, we noticed that the distribution of beam size measurements had a pronounced high-side tail, as may be seen in Fig. \ref{fig:high_side_tail}. To minimize the errors due to this, we sampled the beam size at each knob value several times and averaged measurements near the minimum to determine the actual beam size. Applying this method to CESR with a 5 GeV lattice, we were able to reduce the beam size from 31.2 $\mu m$ $\pm$ 0.3 $\mu m$ to 28.3 $\mu m$ $\pm$ 0.2 $\mu m$. From measurements of the vertical dispersion and coupling near the vbsm, we estimate that this corresponds to a vertical emittance of 31 pm-rad. Given our horizontal emittance of 100 nm-rad, we find that our emittance coupling is 0.03\%.

We wished to see whether or not our knobs are orthogonal in the emittance, so that emittance gains obtained from one knob are not affected by the use of a second knob. To test this, after minimizing the beam size once using the emittance-tuning knobs, we applied the knobs in order a second time. The resulting beam size was 28.7 $\mu m$ $\pm$ 0.3 $\mu m$, similar to the 28.3 $\mu m$ $\pm$ 0.2 $\mu m$ we achieved after one pass. The fact that there was no significant improvement when using the knobs a second time suggests that they are indeed orthogonal in emittance.

In order to verify that we were truly minimizing the emittance, and not simply reducing the local coupling and dispersion at the vbsm source point, which would also bring about a reduction in the size of the beam, we designed knobs which minimize the emittance while leaving the local dispersion and coupling unchanged, as well as while minimizing orbit shifts at the undulator and east collimator, where there are narrow apertures and so a high risk of beam loss. We did this by using the BMAD simulation on the same CESR lattice as we used to get the Hessian matrix to obtain the gradients of the orbit, dispersion, and coupling at the locations where we wanted them to be unchanged. We then removed those magnet combinations from our parameter space before making our Hessian for the vertical emittance. Simulations show that the eigenvectors extracted from this modified Hessian are still effective at reducing the emittance.

Using these new knobs to constrain the beam properties near the vbsm source point on the 5 GeV lattice, we were able to reduce the beam size from 31.2 $\mu m$ $\pm$ 0.3 $\mu m$ to 29.6 $\mu m$ $\pm$ 0.2 $\mu m$. This may be compared with the result of using the unconstrained knobs, which brought the beam size to 28.3 $\mu m$ $\pm$ 0.2 $\mu m$. We see then that most of our observed beam size reduction does indeed stem from a real reduction in the emittance.

\begin{figure}[!htb]
   \centering
   \includegraphics*[width=225pt]{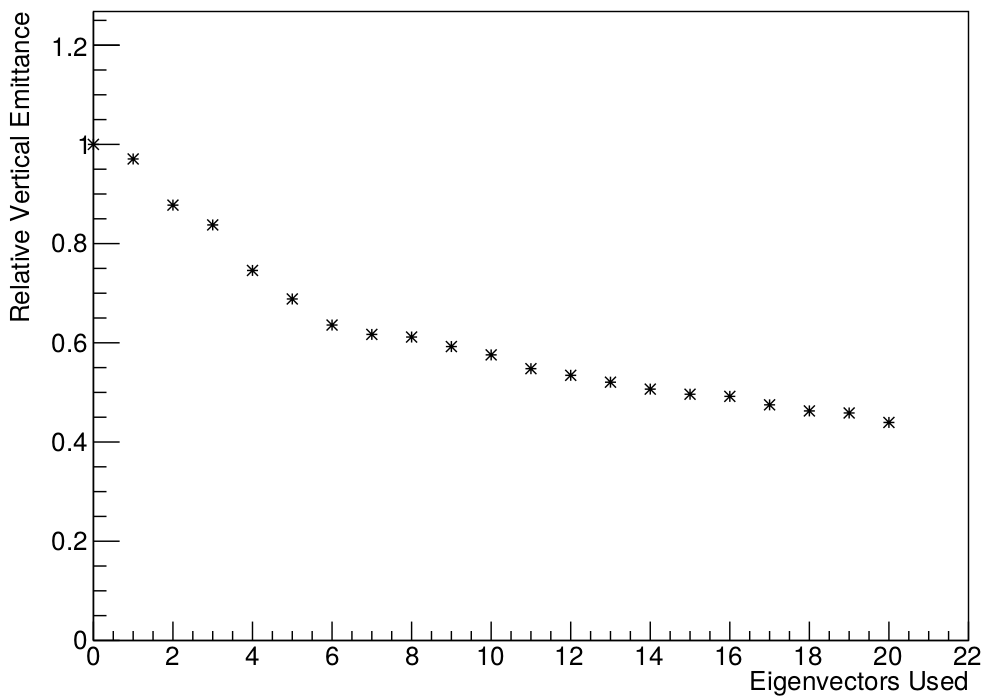}
   \caption{Change in emittance relative to starting emittance in simulation using knobs with only skew quadrupoles. We have turned the knobs individually to minimize beam size, and have averaged the results of 99 versions of the same 5 GeV lattice with random errors.}
   \label{fig:skew_emittance_reduction}
\end{figure}

\begin{figure}[!htb]
   \includegraphics*[width=225pt]{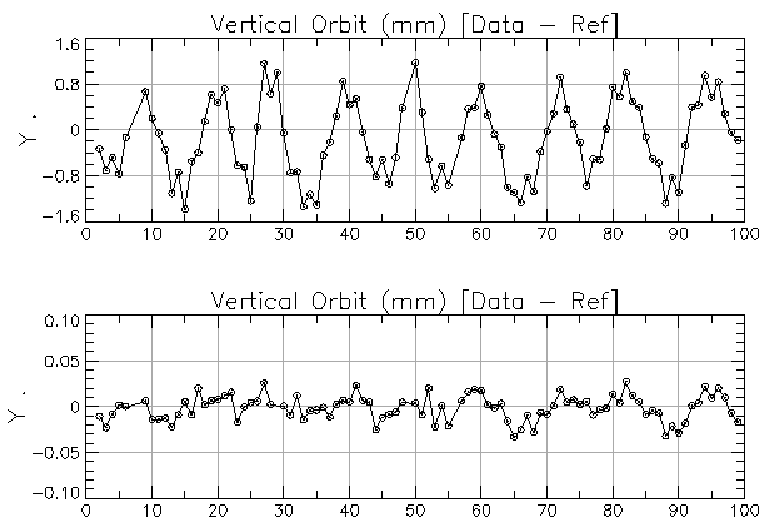}
   \caption{We may compare the orbit shift caused by the knobs obtained from the full Hessian (top) to that caused by the skew-only knobs (bottom). We see that the latter shift the orbit far less than the former; note the difference in scales.}
   \label{fig:side_by_side}
\end{figure}

Prior results had also shown that, if we only used the skew quadrupoles for our emittance-correction and ignored the vertical kickers, we would get virtually no orbit shifts while maintaining much of our ability to fix the emittance.\cite{cite:napac} The fact that use of such knobs does not adversely affect our ability to tune the emittance may be seen by comparing Fig. \ref{fig:emittance_reduction} and Fig. \ref{fig:skew_emittance_reduction}. When we tested these skew-only knobs on CESR, we found that they were able to reduce the beam size to 28.9 $\mu m$ $\pm$ 0.3 $\mu m$, comparable to the 28.3 $\mu m$ $\pm$ 0.2 $\mu m$ we obtained from the unconstrained knobs. Moreover, the orbit shift due to tuning with the skew-only knobs was significantly less than what we saw when we used knobs based on the full Hessian, as may be seen in Fig. \ref{fig:side_by_side}.

\section{Future Work}

So far, we have been successful in preventing changes to beam parameters which we have already optimized. However, we would also like to use this knowledge of which sets of magnets have already been fixed to identify which knobs will no longer contribute significantly to emittance reduction. We have also been restricting ourselves to minimizing using one knob at a time. However, other methods may be more efficient. For example, within the assumption that the knobs are orthogonal and quadratic in the vertical emittance, performing a fit directly to an N-dimensional paraboloid requires fewer fit parameters than fitting N individual parabolas. We are also interested in exploring the applicability of genetic algorithms to this problem, since the reduction in the dimensionality of the parameter space made possible by these knobs will permit more efficient use of such methods.

\section{Conclusions}

We have demonstrated the successful implementation of the theory of sloppy models and the RCDS algorithm to reduce the emittance at the Cornell Electron Storage Ring, and have shown that it permits additional improvements beyond our best previous method. Moreover, we have shown that by making judicious restrictions on our parameter space, we are able to prevent certain desirable beam properties from being affected by this minimization process without loss to our ability to reduce the vertical emittance.

\section{Acknowledgments}

We would like to thank Suntao Wang for his assistance with the vbsm, James Shanks for his aid with the simulations, and Danilo Liarte for his help with the theoretical underpinnings of this project. This work was supported by the Department of Energy under grant number DE-SC$ $0013571. W.F.B. would also like to acknowledge the support of the National Science Foundation Graduate Research Fellowship Program under grant number DGE-1144153.


\null  

\end{document}